\documentclass[prl,twocolumn]{revtex4-1}
\usepackage[a4paper,top=1.50cm,bottom=1.50cm,left=1.50cm,right=1.50cm]{geometry}
\usepackage{float}
\usepackage{amsmath}
\usepackage{amsfonts}
\usepackage{amssymb}
\usepackage{esint}
\usepackage{mathtools}
\usepackage{url}
\usepackage[pdftex]{hyperref}
\usepackage{bm}
\usepackage{graphicx}
\usepackage{color}
\usepackage{tikz}
\DeclareMathOperator{\tr}{tr}
\def\la{\left\langle}
\def\ra{\right\rangle}
\newcommand{\nn}{\nonumber}
\def\g{\widehat{\mathfrak{G}}}
\def\mG{\mathfrak{G}}
\def\mF{\mathfrak{F}}
\renewcommand{\t}[1]{\widehat\tau_{#1}}
\def\e{\varepsilon}
\def\ps{\mathbf{p}_s}
\def\js{\mathbf{j}_s}
\def\vB{\mathbf{B}}
\def\vA{\mathbf{A}}
\def\vH{\mathbf{H}}
\def\vz{\mathbf{z}}
\def\vy{\mathbf{y}}
\def\vv{\mathbf{v}}
\def\vp{\mathbf{p}}
\def\vj{\mathbf{j}}

\def\p{\mathbf{\hat p}}

\def\ns{\negmedspace}
\def\nicefrac#1#2{\genfrac{}{}{}{1}{#1}{#2}}
\renewcommand{\Im}{\mathsf{Im}}
\begin{document}
\title{
The Effect of Inhomogeneous Surface Disorder on the Superheating Field of 
Superconducting RF Cavities 
}
\author{Vudtiwat~Ngampruetikorn}
\email{wave@northwestern.edu}
\author{J.~A. Sauls}
\email{sauls@northwestern.edu}
\affiliation{
Center for Applied Physics \& Superconducting Technologies 
\\
Department of Physics \& Astronomy, Northwestern University, Evanston, IL 60208
\\
Fermi National Accelerator Laboratory, Batavia IL, 60510-5011
}
\date{\today}

\begin{abstract}
Recent advances in surface treatments of Niobium superconducting radio frequency (SRF) cavities
have led to substantially increased Q-factors and maximum surface field. This poses theoretical 
challenges to identify the mechanisms responsible for such performance enhancements.
We report theoretical results for the effects of inhomogeneous surface disorder on the 
superheating field --- the surface magnetic field above which the Meissner state is 
globally unstable. 
We find that inhomogeneous disorder, such as that introduced by infusion of Nitrogen into 
the surface layers of Niobium SRF cavities, can increase the superheating field above the maximum 
for superconductors in the clean limit or with homogeneously distributed disorder. 
Homogeneous disorder increases the penetration of screening current, but also suppresses the 
maximum supercurrent.
Inhomogeneous disorder in the form of an impurity diffusion layer biases this trade-off by
increasing the penetration of the screening currents into cleaner regions with larger 
critical currents, thus limiting the suppression of the screening current to a thin dirty 
region close to the surface. 
Our results suggest that the impurity diffusion layers play a role in enhancing 
the maximum accelerating gradient of Nitrogen treated Niobium SRF cavities.
\end{abstract}
\maketitle

\emph{Introduction} --- Type-II superconductors admit two thermodynamic phases in the presence 
of an external magnetic field $H$. The Meissner state is the equilibrium state for fields below 
a lower critical field $H<H_{c_1}$, while the Abrikosov state, characterized by the penetration of
quantized flux into the bulk of the superconductor, is the thermodynamically stable phase for 
fields $H_{c_1}<H<H_{c_2}$, where $H_{c_2}$ is the critical field above which the superconductor 
becomes normal for any temperature $T\le T_c$.  
Superconductors in the Meissner state exhibit perfect diamagnetism by generating an internal field,
that exactly screens the external field. The source of the screening field is a dissipationless 
supercurrent, ``screening current'', confined to the vacuum-superconductor interface. The 
screening current  penetrates into the superconductor over a mesoscopic length scale, the London 
penetration depth $\lambda_L$, which is sensitive to disorder. 
The magnitude of the screening current increases linearly with the applied field until
the cost in kinetic energy of maintaining perfect diamagnetism is outweighed by the reduction in
Gibbs energy via flux penetration into into the bulk. For type-II superconductors flux is 
quantized in units of $\Phi_{0}=hc/2e$ and confined in tubes of radius of order the London 
penetration depth, and the lower critical field for flux penetration is 
$H_{c_1}=\Phi_0/2\pi\lambda_L^2$, which for SRF grade Nb is typically of order 
$H_{c_1}\approx 30\,\mbox{mT}$, or an accelerating field of
$E_{\text{ac}}\approx 25\,\mbox{MV/m}$.
 
Above $H_{c_1}$ the Abrikosov state, with an array of quantized flux lines, is the 
thermodynamically stable phase. Motion of quantized flux generates Joule losses and is 
detrimental to to the performance of SRF cavities for particle acceleration.
Understanding, and thus engineering, materials properties and physical processes governing the 
breakdown of the Meissner state is crucial for developing strategies to improve the performance of 
SRF cavities.

One key feature is that Meissner state can be maintained for fields higher than $H_{c_1}$ as a 
meta-stable phase, made possible by a surface energy barrier to flux penetration~\cite{bea64}. At 
sufficiently high field, the so-called superheating field, $H_\text{sh}>H_{c_1}$, the surface 
barrier vanishes, and quantized flux lines proliferate leading to dissipation 
under RF excitation. 

The superheating field depends on the geometry of the vacuum-superconductor interface as well as the
spatial distribution of disorder within the region of the screening currents.
For a planar half-space geometry the effects of homogeneous disorder and engineered multilayer 
superconductor-insulator structures has been studied~\cite{lin12,gur15,lia16,lia17,kub17}. The 
main results are that homogeneous disorder increases the penetration depth, but reduces the 
critical current, with a modest enhancement of the superheating field at low 
temperatures~\cite{lin12}. The superheating field may be increased by introducing insulating 
layers to retard flux line penetration. 
Here we report results of a theoretical investigation of the effects of an impurity diffusion 
layer, i.e. a smoothly varying, coarse-grained impurity density within the region of the 
screening currents, on SRF cavities such as Nitrogen infused into Niobium~\cite{gra17}.

In general it is technically challenging to obtain quantitative predictions for the superheating 
field as one must consider the stability of the Meissner state to inhomogeneous fluctuations of 
order parameter and charge currents, as well as nucleation of vortices around impurities, 
inclusions or sharp structures at the vacuum-superconductor interface.
Here we consider the upper limit for the superheating field, which is the lowest surface field at 
which the supercurrent density reaches the local critical current density at some point within the 
screening region near the vacuum-superconductor interface.
This condition provides an upper bound to the superheating field since any increase in the local 
condensate momentum - equivalently the local vector potential - cannot increase the supercurrent 
density. At the superheating field the Meissner state is unstable to arbitrarily small 
perturbations of the order parameter and electromagnetic (EM) field. For extreme type-II 
superconductors, this approach is equivalent to the stability condition with respect to inhomogeneous
fluctuations of order parameter and the associated EM response~\cite{cat08}.

Type-II superconductors are characterized by Ginzburg-Landau (GL) parameter, 
$\kappa=\lambda_L/\xi \ge 1/\sqrt{2}$, where $\lambda_L$ denotes the London penetration depth 
and $\xi$ is the superconducting coherence length. 
Pure Nb is weakly type II with $\kappa\approx 1$. However, disorder leads to increased field 
penetration with $\kappa\gg 1$ in the ``dirty limit'', $\hbar/\tau\gg\Delta$, where $\tau$
is the mean quasiparticle-impurity collision time.
In this strong type-II limit quasiparticles and Cooper pairs respond locally to a nearly 
uniform EM field. 

Here we consider superconductors in the strong type-II limit occupying the half space $x>0$ in 
the presence of an external magnetic field, $\vH_\text{a}=H_a\hat\vz$, applied parallel to the 
vacuum-superconductor interface. We include the effects of an impurity diffusion layer on the 
current response into the quasiparticle-impurity scattering rate and pairing self-energy. 
Based on Eilenberger's quasiclassical transport theory~\cite{eil68} we compute the superfluid 
momentum $\ps=p_s(x)\hat\vy$, screening supercurrent $\js=j_s(x)\hat\vy$ and local magnetic 
induction $\vB=B(x)\hat\vz$ self-consistently.
The superheating field $H_\text{sh}$ is the value of the surface field, $B(0)$, at which the 
supercurrent density reaches the local critical value anywhere in the screening region of the 
superconductor, i.e. $\min_x[j_c(x)-|j_s(x)|]=0$. Note that the critical current density 
$j_c(x)$ is a function of position due to the inhomogeneous impurity diffusion layer. 

\emph{Methods} ---
For a superconductor in the strong type-II limit, with an impurity diffusion layer that also 
varies on a length scale much longer than $\xi$, we develop the Eilenberger transport equation
as a perturbation expansion in the small ratios, $\epsilon\in\{\xi/\lambda_L,\,\xi/\zeta\}$, 
where $\zeta$ is the characteristic penetration length of the impurity diffusion 
layer (Appendix).
To leading order in $\epsilon$ the current response is determined by the retarded quasiclassical 
propagator obtained from the homogeneous solution of the quasiclassical transport equation, 
but evaluated with the Doppler shifted excitation spectrum determined by the local condensate 
momentum, $p_s(x)$, and the local impurity self-energies, $\Sigma_{\text{imp}}(x)$ and 
$\Delta_{\text{imp}}(x)$,
\begin{align}
\g(\p,\e,x)
&\ns=\ns-\pi
\frac{[\tilde\e(\e,x)\ns-\ns\vv_f\cdot\ps(x)]\t3\ns-\ns{\tilde\Delta}(\e,x)(i\sigma_y\t1)}
{\sqrt{|\tilde\Delta(\e,x)|^2-[\tilde\e(\e,x)-\mathbf v_f\cdot\ps(x)]^2}}
\nn
\\
&\equiv-\pi[\mG(\p,\e,x)\t3-\mF(\p,\e,x) (i\sigma_y\t1)],
\label{eq:propagator}
\end{align}
where $\t i$ and $\sigma_i$ denote the Pauli matrices in particle-hole and spin space, 
respectively, $\p$ is the direction defined by a point on the Fermi surface, $\vp=p_f\p$, and 
$\vv_f =v_f\p$ is the corresponding Fermi velocity. 
In the absence of vortices the superfluid momentum can be related to the vector potential 
via $\vp_s=(-e/c)\vA$, where we have fixed the gauge by absorbing the gradient of the phase
of the condensate into $\vA$.
The diagonal and off-diagonal propagators, $\mG$ and $\mF$, encode the information about 
the local equilibrium quasiparticle and Cooper pair spectral functions.

The impurity renormalized quasiparticle excitation energy and off-diagonal pairing energy
can then be expressed as
\begin{equation}
\begin{aligned}
\tilde\e(\e,x)	
&=\e+\gamma(x)\la \mG(\p,\e,x)\ra_\p	
\,,
\\
\tilde\Delta(\e,x)
&=\Delta(x)+\gamma(x)\la \mF(\p,\e,x)\ra_\p
\,,
\end{aligned}
\end{equation}
$\la\dots\ra_\p$ denotes an angular average over the Fermi surface and $\gamma(x)$ is 
the local impurity scattering rate.
The order parameter, $\Delta(x)$, satisfies the mean-field BCS gap equation, 
\begin{equation}
\label{eq:Delta_mf}
\Delta(x)=\frac{g}{2}\fint d\e \tanh\frac{\e}{2T}\Im\la f(\p,\e,x) \ra_\p,
\end{equation}
where $g$ is the pairing interaction, and the integration is extends
over the low-energy bandwidth set by the Debye energy.
The set of equations for the propagators, self energies, and mean-field gap equation
are derived in the Appendix, including the next-to-leading order corrections 
from gradients of the leading order local propagators, which are smaller by a factor of 
order $\epsilon$.

\begin{figure}
\centering
\includegraphics[width=8.3cm]{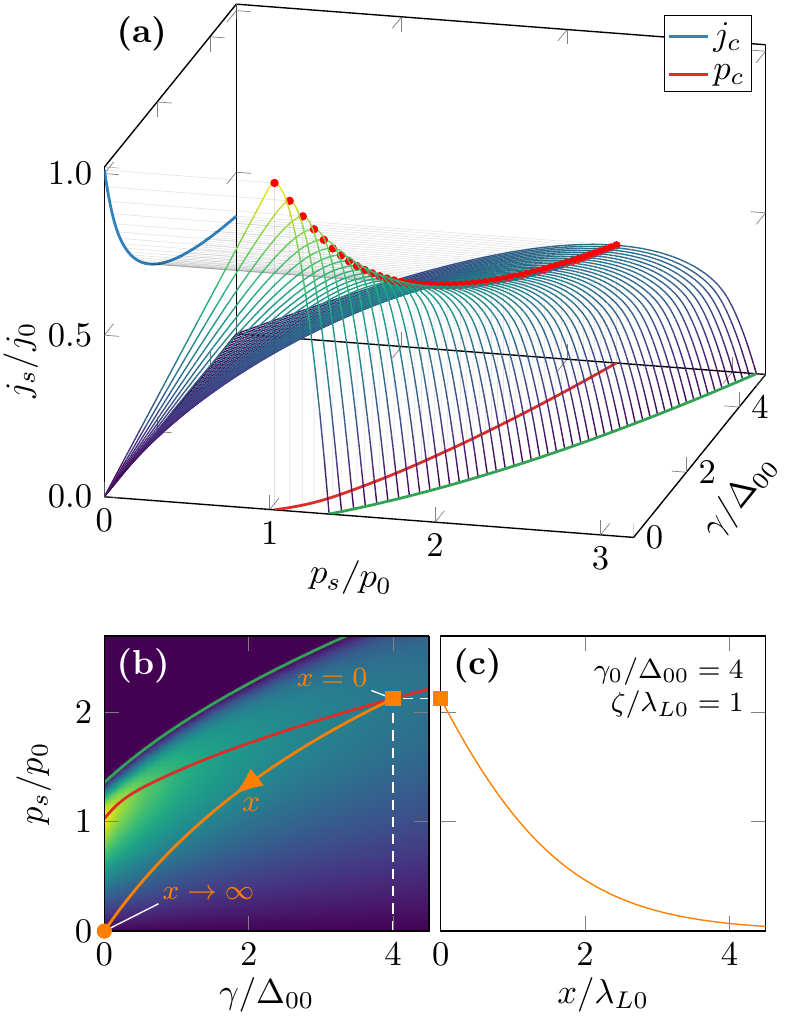}
\caption{
\label{fig:js}
\textbf{Panel~(a)}: the supercurrent density $j_s$ as a function of condensate momentum 
$p_s$ and impurity scattering rate $\gamma$ at $T=0$. For fixed $\gamma$, the critical 
current, $j_c$, and condensate momentum, $p_c$, correspond to the values at which $j_s$ 
is maximum (red filled circles). 
The critical current, $j_c$, decreases with increasing $\gamma$ (blue line), whereas the 
critical condensate momentum, $p_c$, increases with increasing $\gamma$ (red line). 
For $p_s>p_c$, the Meissner current is unstable.
\textbf{Panel~(b)}: a typical solution of Eq.~\eqref{eq:maxwell} at the superheating 
field (orange curve), overlaid on a color density plot of $j_s$ (same color scale as in (a)).
The boundary conditions at $x=0$ and $x\to\infty$ are set by the superheating condition, 
$j_s(0)=j_c(0)$ (orange square) and the Meissner condition, $p_s(\infty)=0$ (orange circle), 
respectively. 
The impurity scattering rate varies with $x$ (arrow) based on Eq.~\eqref{eq:gamma} and 
vanishes as $x\to\infty$ (orange circle).
\textbf{Panel~(c)}: spatial profile of the condensate momentum for the case shown in 
panel (b) with $\gamma_0/\Delta_{00}=4$ and $\zeta/\lambda_{L0}=1$.
}
\end{figure}

The solution for the field penetration into the inhomogeneous Meissner region of the 
superconductor is obtained from the local current response, which is in general a nonlinear 
function of the condensate momentum, $\ps(x)$, combined with Amp\`ere's equation. 
The latter equation can be expressed as
\begin{equation}
\label{eq:maxwell}
\partial_x^2\ps(x)-\frac{4\pi e}{c^2}\vj_s[\ps(x),\gamma(x)] = 0
\,,
\end{equation}
where the supercurrent is obtained from the local solution for the quasiclassical 
propagator,
\begin{equation}
\label{eq:j_s}
\vj_s(x)=-eN_f \int d\e\tanh\frac{\e}{2T}\la \vv_f\,\mathcal A(\p,\e,x)\ra_\p
\,,
\end{equation}
where $N_f = p_f^2/2\pi^2\hbar^3 v_f$ is the normal-state density of states, per spin, 
at the Fermi level.
The Meissner current sums the charge current contributions from the states comprising both
the negative energy condensate, as well as thermally excited Bogoliubov quasiparticles,
governed by the angle-resolved spectral function, 
${\mathcal A}(\p,\varepsilon;x) \equiv \nicefrac{-1}{\pi}\Im\,\mG(\p,\e;x)$,
and the thermal distribution function, $\Phi(\e)=\tanh(\e/2T)$.

To determine the magnetic field distribution in the superconductor, we find the self-consistent
condensate momentum distribution, $p_s(x)$, that determines the supercurrent, $j_s(x)$, given
by Eq. \eqref{eq:j_s}, and is also the solution of Amp\`ere's law given by Eq. \eqref{eq:maxwell}.
Amp\`ere's law is also supplemented by boundary conditions at the surface and the asymptotic
condition far from the vacuum-superconductor interface,
\begin{equation}\label{eq:boundary_cond}
\vspace*{-2mm}
\nabla\times\mathbf p_s(x)|_{x=0}= (-e/c)\mathbf H_\text{a}
,\,\text{and}\,
\lim_{x\to\infty}\mathbf p_s(x)=0
\,.
\end{equation}
The asymptotic condition reflects the fact that the Meissner state exhibits perfect 
diamagnetism. Equations 
\eqref{eq:propagator}-\eqref{eq:boundary_cond} 
constitute a closed 
set of equations which are solved self-consistently.
The local magnetic induction can then be computed directly from 
$\vB(x)=(-c/e)\partial_x p_s(x)\,\hat\vz$. 

In order to determine the superheating field we first solve 
Eqs. \eqref{eq:propagator}-\eqref{eq:boundary_cond} self-consistently for fixed temperature,
$T$, external field, $H_a$, and impurity distribution, $\gamma(x)$, which yields
the self-consistently determined spatial profiles for the condensate momentum, $p_s(x)$,
and Meissner screening current, $j_s(x)$. 
The spatial profile of the magnetic field is obtained from the condensate momentum 
$B(x)=(-c/e)\partial_x p_s(x)$.
To determine the superheating field, we determine the surface field, $B(0)=H_{\text{sh}}$,
at which the supercurrent and the superfluid momentum reach local critical values anywhere in 
the Meissner screening region.

\emph{Impurity Diffusion Layer} --- For concreteness we model the impurity diffusion layer 
as exponential decay from the vacuum-superconducting interface, 
$n_{\text{imp}}(x) = n_0\,\exp{(-x/\zeta)}$; or equivalently a local scattering 
rate of the form,
\begin{equation}\label{eq:gamma}
\gamma(x) = \gamma_0 e^{-x/\zeta},
\end{equation}
where $\gamma_0$ denotes the impurity scattering rate at $x=0$ and $\zeta$ is the impurity 
diffusion length. Similar results are obtained based on a Gaussian diffusion layer.
This model qualitatively captures the impurity distribution in Nitrogen treated SRF cavities,
i.e. high impurity concentration near the surface and very low impurity concentration in the 
bulk~\cite{gra17}. 
We confine our analysis to diffusion lengths that are large compared to 
the coherence length, $\zeta\gg\xi$, so that we can  evaluate the propagator with the 
locally homogeneous solution in Eq.~\eqref{eq:propagator}.
In this model the condensate momentum first reaches the critical value at the surface, i.e., the 
superheating condition is given by $p_s(0)=p_c(0)$, where $p_c(0)$ is the critical condensate 
momentum determined by the maximum scattering rate, $\gamma_0$.

\begin{figure}
\centering
\includegraphics[width=7.5cm]{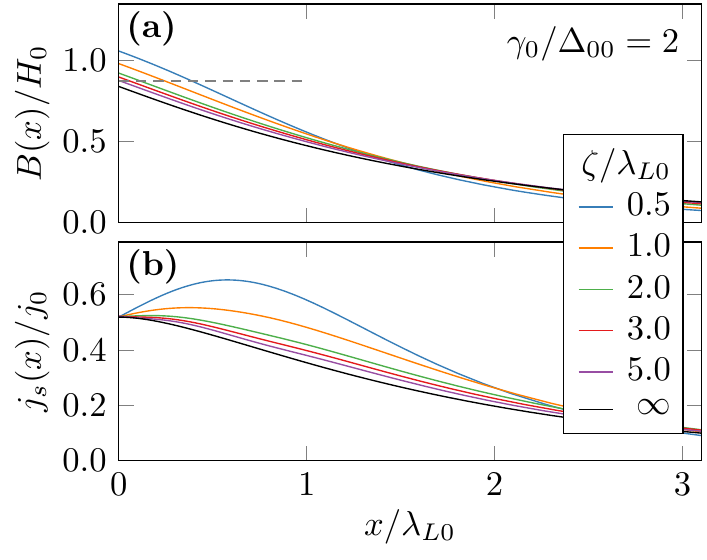}
\caption{
\label{fig:imp_length_effect}
The magnetic field and current density profiles, $B(x)$ and $j_s(x)$, at the superheating 
field for temperature $T=0$, various impurity diffusion lengths $\zeta$ shown in the legend, 
and surface scattering rate $\gamma_0/\Delta_{00}=2$, where $\Delta_{00} = 1.78\,T_c$ is the 
zero-temperature BCS gap in the clean limit. 
\textbf{Panel~(a)}: $B(x)$ in units of the zero-temperature, clean limit critical field,
$H_0=\sqrt{4\pi N_f \Delta_{00}^2}$. The superheating field $H_\text{sh}=B(0)$ increases with 
decreasing diffusion length $\zeta$, and exceeds the superheating field for the case of 
homogeneous disorder with scattering rate $\gamma_0$ (dashed line).
\textbf{Panel~(b)}: $j_s(x)$ in units of the zero-temperature, clean-limit critical current, 
$j_0 = e\,n\,\Delta_{00}/p_f$.
The current density builds up away from the surface as $\zeta$ decreases, leading to larger 
total screening currents (the area under the curves) and thus higher superheating 
fields.
}
\end{figure}

\emph{Results} --- Figure \ref{fig:imp_length_effect} shows the magnetic field and current density 
profiles at the superheating field for a scattering rate at the surface, $\gamma_0/\Delta_{00}=2$, 
where $\Delta_{00}$ is the excitation gap at $T=0$ in the clean limit. 
We present results for impurity diffusion lengths ranging from the homogeneous limit, 
$\zeta\to\infty$, to $\zeta/\lambda_{L0} = 0.5$, scaled in units of the clean-limit, $T=0$, 
zero-field London penetration depth, $\lambda_{L0}=1/(8\pi e^2 v_f^2 N_f/3c^2)^{\nicefrac{1}{2}}$,
but restricted to $\zeta\gg\xi_0$. 
Fig.~\ref{fig:imp_length_effect}(a) shows that the superheating field, given by the field at 
$x=0$, increases with decreasing impurity diffusion length, and exceeds the absolute maximum 
superheating field of $H_{\text{sh}}^{\infty} \approx 0.88\,H_0$ reported in Ref.~\cite{lin12} 
for homogeneous disorder with $\gamma_0/\Delta = 0.3$ (shown as the dashed line).
Our analysis also confirms the prediction of Ref.~\cite{lin12} for the effect of 
homogeneous disorder.

To understand how an inhomogeneous impurity distribution leads to an increase in the 
superheating field consider the current density profiles shown in 
Fig.~\ref{fig:imp_length_effect}(b). 
At the superheating field the current density at $x=0$ is equal to the local critical current 
density, which is determined by $\gamma_0$ in each case. 
However, away from the surface a shorter impurity diffusion length results in a reduced impurity 
density and therefore larger current density for a given value the local condensate momentum. 
Indeed for sufficiently short impurity diffusion lengths the current density peaks 
at a finite distance from the vacuum-superconductor interface, resulting in a larger integrated
screening current, $J=\int_0^\infty dx\, j(x)$, more effective screening of the field, 
and thus a higher superheating field.

\begin{figure}
\centering
\includegraphics[width=7.5cm]{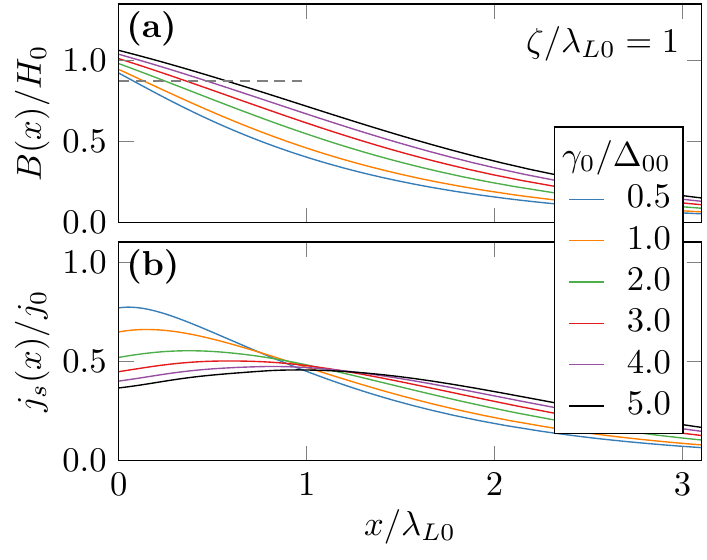}
\caption{
Similar plots as Fig.~\ref{fig:imp_length_effect}, but for a fixed impurity diffusion length 
of $\zeta/\lambda_{L0}=1$ as a function of surface scattering rate, $\gamma_0$, shown in the
legend. The superheating field, $H_\text{sh}=B(0)$, exceeds the theoretical maximum for the 
case of homogeneous disorder~\cite{lin12} (dashed line) over the whole range of $\gamma_0$.
}
\label{fig:scatt_rate_effect}
\end{figure}

Figure~\ref{fig:scatt_rate_effect} shows the magnetic field and current density profiles at 
the superheating field for a fixed impurity diffusion length $\zeta/\lambda_{L0}=1$, for 
a range of maximum impurity scattering rates $\gamma_0$. In Fig.~\ref{fig:scatt_rate_effect}(a) 
the magnetic field penetrates deeper into the superconductor with increasing impurity scattering
at the surface, \emph{and} the superheating field increases above the absolute maximum 
superheating field for homogeneous disorder~\cite{lin12} (dashed line). The screening current 
penetrates deeper further into the superconductor, but is suppressed for $x\lesssim\lambda_{L0}$,
as shown in Fig.~\ref{fig:scatt_rate_effect}(b). 
However, the local suppression of the current near the surface is overcompensated by the 
increase in the screening current for $x\gtrsim\lambda_{L0}$ over a longer
effective penetration depth, leading to an increase in the superheating field.

\begin{figure}
\centering
\begin{tikzpicture}
	\node[inner sep=0pt] (Bsh) at (0,0) {\includegraphics[width=7.5cm]{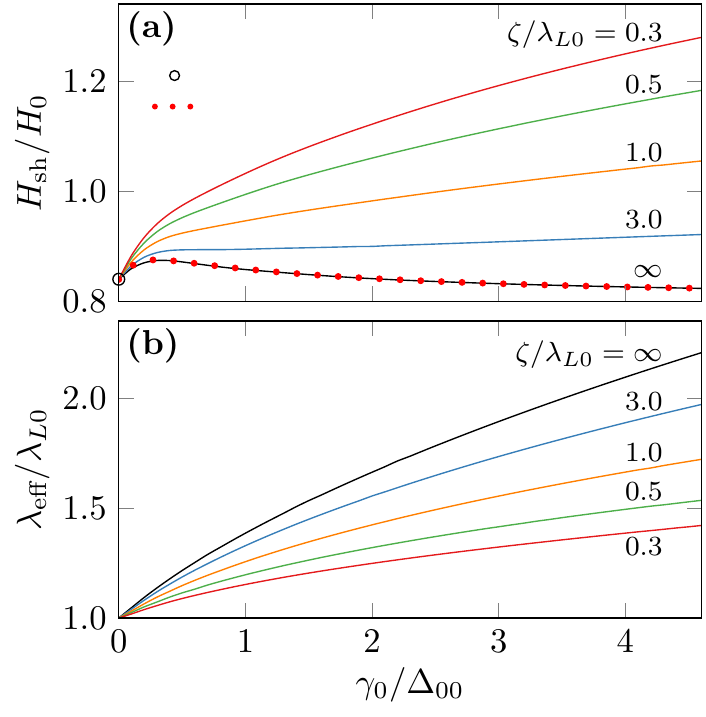}};
	\node[anchor=north west,shift={(20.5mm,-9.1mm)}] 
			at (Bsh.north west) {Ref.~\cite{lin12}};
	\node[anchor=north west,shift={(20.5mm,-5.6mm)}] 
			at (Bsh.north west) {Refs.~\cite{gal66,cat08}};
\end{tikzpicture}
\caption{
The superheating field, $H_{\text{sh}}$, and effective penetration depth, $\lambda_\text{eff}$, 
in superconductors with an impurity diffusion layer [Eq.~\eqref{eq:gamma}] as functions of the 
surface scattering rate, $\gamma_0$, for various impurity 
diffusion lengths, $\zeta$, shown in the legend. 
We compare our results for the superheating field with previous calculations in the clean 
limit~\cite{gal66,cat08} (black open circle) and for homogeneous disorder~\cite{lin12} 
(red circles).
}
\label{fig:Bsh}
\end{figure}

Figure~\ref{fig:Bsh} summarizes our results for the superheating field in impurity diffusion 
layers at $T=0$. Disorder affects the superheating field via two competing mechanisms. 
First, the effective penetration depth, defined as 
\begin{equation}
\lambda_\text{eff}\equiv B(0)^{-1}\,\int_0^\infty dx\,B(x)
\,,
\end{equation}
increases with disorder as shown in Fig.~\ref{fig:Bsh}(b). As a result, the screening current 
penetrates deeper into the superconductor, increasing the total screening current, and as a 
result the superheating field.
Second, impurity scattering suppresses the critical current and superheating field, c.f.
Fig.~\ref{fig:js}(a). For homogeneous disorder the increase in the effective penetration depth 
is dominant at low scattering rates, while the suppression of supercurrent dominates at higher 
scattering rates. As a result the superheating field develops a peak at a relatively modest
level of disorder, $\gamma_0/\Delta_{00}\approx 0.3$ shown in Fig.~\ref{fig:Bsh}(a) for 
$\zeta/\lambda_{L0}=\infty$, with $H_{\text{sh}}\approx 0.87\,H_0$.
However, in impurity diffusion layers the suppression of supercurrent is confined to the 
region near the surface $x\lesssim\zeta$, while due to a longer effective penetration depth
[c.f. Fig.~\ref{fig:scatt_rate_effect}(b)],
the screening current shifts to the relatively clean region with $x\gtrsim\zeta$.
This results in a superheating field that increases with the surface scattering rate, 
as shown in Fig.~\ref{fig:Bsh}(a) for diffusion lengths $\zeta/\lambda_{L0}\gtrsim 3.0$.

We note that our analysis for the superheating field based on the local critical depairing 
current is equivalent to the stability analysis of the Meissner state 
presented in Refs.~\cite{gal66,cat08,lin12}. 
Indeed our results agree with the previous calculations based on analyses of the 
thermodynamic potential.
In particular we obtain $H_\text{sh}/H_0\approx0.84$ for clean type-II superconductors
as reported in Refs.~\cite{gal66,cat08,lin12} (black open circle in Fig. \ref{fig:Bsh}(a)).
Our results also agree with those of Ref.~\cite{lin12} for the limit of homogeneous disorder, 
shown as the red data points in Fig.~\ref{fig:Bsh}(a).

So far we have considered the extreme type-II limit with $\kappa^{-1}=\xi/\lambda_{L}\to 0$. 
Niobium, the material of choice for SRF applications, is marginally type II in the clean
limit with $\kappa\approx1$ \cite{str62,max65,fin66}. 
Cavity-grade Niobium has surface disorder, and is treated with Nitrogen impurities to increase 
performance, both of which increase the GL parameter, thus suppressing the corrections to our 
theory which are of order $\kappa^{-2}$, as shown in the Appendix.
Thus, we believe this work provides new insight into the role of inhomogeneous disorder on 
the superheating field in Nitrogen-infused Niobium SRF cavities. 
Moreover, our results have implications for the other superconducting materials considered for 
SRF applications, such as Nb$_3$Sn and MgB$_2$, both of which are strong type-II superconductors 
with $\kappa\gtrsim20$~\cite{orl79,jin02}.

\emph{Summary and Outlook} --- We report a theoretical investigation based on microscopic theory of
inhomogeneous superconductors of the effects of impurity diffusion layers on the superheating 
field of superconducting RF cavities, the limiting magnetic field beyond which the Meissner 
state is unstable. 
A key result is that the introduction of an impurity diffusion layer, for example by Nitrogen 
infusion into Niobium, can increase the superheating field of SRF cavities above the  
maximum allowed superheating field predicted for the homogeneous disorder 
model~\cite{lin12}.
The underlying mechanism is the increase in screening current resulting from increased field 
penetration depth which overcompensates suppression of Meissner current in the relatively 
thin dirty region near the surface. 
Our results strongly suggest that the impurity diffusion layers play a role in 
enhancing the maximum accelerating gradient of treated SRF cavities.
Although the increase in the superheating field appears to be generic to impurity diffusion 
layers, the magnitude of the increase depends on specific impurity profiles, suggesting that it 
might be possible to further increase the superheating field by engineering disorder profiles. 

{\it Acknowledgments --}
We thank Anna Grassellino, Alex Romanenko, Mattia Checchin and Martina Martinello for discussions 
on N-doped Niobium SRF cavities which motivated this research.
The research of the authors is supported by National Science Foundation Grant PHY-1734332 and 
the Northwestern-Fermilab Center for Applied Physics and Superconducting Technologies.
This work was finalized at the Aspen Center for Physics, which is supported by National Science Foundation grant PHY-1607611.

\begin{widetext}
\subsection{Appendix: Long wavelength expansion of the Eilenberger Transport Equations}

We solve Eilenberger's transport equation as an expansion in the ratio of length 
scales, $\epsilon=\{\xi/\lambda_{L},\xi/\zeta\}$, for the Meissner state of inhomogeneous type-II 
superconductors with an impurity diffusion layer. The propagator to zeroth order in $\epsilon$ is
the local solution given in Eq.~\eqref{eq:propagator}.
We also show that the leading order corrections to our calculations for the 
superheating field are smaller by a factor of $\epsilon^2\sim\kappa^{-2}$.

We take the superfluid momentum to be along the $y$ axis. The quasiclassical transport 
equation is then~\cite{eil68},
\begin{equation}
\left[\left(\e-p_s(x)v_y\right)\t3-\widehat\Sigma(\e,x),\g(\p,\e,x)\right]
+ 
i \hbar v_x\partial_x\g(\p,\e,x)=0
\,,
\end{equation}
where $v_x=v_f\hat p_x$, $v_y=v_f\hat p_y$.  
Spatial dependences enter via the condensate momentum, $p_s(x)$, and the impurity self energy,
$\widehat\Sigma(\varepsilon,x)$, which vary on the characteristic length scales, $\lambda_{L}$ 
and $\zeta$, respectively.
We consider $\zeta\sim\lambda_{L}\ll \xi$, where the coherence length is 
$\xi=\hbar v_f/(2\pi T_{c})$, and introduce the dimensionless distance, 
$s=x/\lambda_{L}$, in which case derivatives of order $\partial_s\sim O(1)$.  
The scaled transport equation becomes,
\begin{equation}
(2\pi T_{c})^{-1}\left[\left(\e- v_f\hat 
p_yp_s(s)\right)\t3-\widehat\Sigma(\e,s),\g(\p,\e,s)\right]
+
i\kappa^{-1}\hat p_x\partial_s\g(\p,\e,s)=0
\,.
\label{eq:transport-scaled}
\end{equation}
The terms defined by the commutator on the \textit{l.h.s.} of 
Eq.~\eqref{eq:transport-scaled} are $\sim O(1)$ since 
$T_{c}$ is the characteristic energy scale in the superconducting state. 
However, the gradient term is proportional to $\kappa^{-1}$, and thus of $O(\epsilon)$. 

We now expand the propagator ($\widehat o\to\g$) and self-energy ($\widehat o\to\widehat\Sigma$)
in the small expansion parameter $\epsilon$,
\begin{equation}
\widehat o=\widehat o^{(0)}+\epsilon\,\widehat o^{(1)} + \epsilon^{2}\, \widehat o^{(2)} + \dots,
\end{equation}
such that terms $\g^{(i)}\sim O(1)$ and $\widehat\Sigma^{(i)}\sim O(T_c)$. 
The zeroth-order terms define the locally homogeneous equation,
\begin{equation}
\left[\left(\e- v_f\hat p_yp_s(s)\right)\t3-\widehat\Sigma^{(0)}(\e,s),\g^{(0)}(\p,\e,s)\right]=0
\,.
\end{equation}
When combined with the Eilenberger's normalization condition, $(\g^{(0)})^2=-\pi^2\widehat 1$, 
we obtain the locally homogeneous propagator in Eq.~\eqref{eq:propagator}. 

The first-order correction to the propagator satisfies
\begin{equation}
(2\pi T_{c})^{-1}
\left\{
\left[\left(\e- v_f\hat p_yp_s(s)\right)\t3-\widehat\Sigma^{(0)}(\e,s),\g^{(1)}(\p,\e,s)\right]
+
\left[\g^{(0)}(\p,\e,s),\widehat\Sigma^{(1)}(\e,s)\right]
\right\}
+i\hat p_x\partial_s\g^{(0)}(\p,\e,s) =0
\,.
\end{equation}
By separating the terms according to their symmetry under $\hat p_x\to-\hat p_x$ and noting 
that both $\g^{(0)}$ and $\widehat\Sigma^{(0)}$ are even in $\hat p_x$, we see that the source 
term generates $\g^{(1)}$ which is odd in $\hat p_x$. Consequently both pairing and ($s$-wave) 
impurity self-energies vanish since they are linear in $\langle \g^{(1)} \rangle=0$. 
Using Eq.~\eqref{eq:propagator}, we eliminate $\widehat\Sigma^{(0)}$ in favor 
of $\widehat \mG^{(0)}$ to obtain, 
\begin{equation}
(2\pi T_{c})^{-1}(-\pi)^{-1}C^{(0)}(\p,\e,s)\left[\g^{(0)}(\p,\e,s),\g^{(1)}(\p,\e,s)\right]
+i\hat p_x\partial_s\g^{(0)}(\p,\e,s)=0,
\end{equation}
where $C^{(0)}$ is a scalar. 
Inverting this equation, we have
\begin{equation}
(-2\pi^2)(2\pi T_{c})^{-1}(-\pi)^{-1}C^{(0)}(\p,\e,s)\g^{(1)}(\p,\e,s)
=-i\hat p_x\g^{(0)}(\p,\e,s)\partial_s\g^{(0)}(\p,\e,s).
\end{equation}
In deriving the above equation we make use of the normalization conditions: 
$(\widehat\mG^{(0)})^2=-\pi^2\widehat1$ and $\widehat \mG^{(0)}\widehat\mG^{(1)}
+\widehat\mG^{(1)}\widehat\mG^{(0)}=0$. 
Finally, we show that $\g^{(1)}$ is purely off-diagonal. 
To see this we note that 
$\g^{(0)}=a_3\t3+a_1(i\sigma_y\t1)$ and thus 
$\g^{(1)}\propto\g^{(0)}\partial_s\g^{(0)}=b_0\hat1+b_2(i\sigma_y\t2)$. In addition since 
$(\g^{(0)})^2=-\pi^2\hat1$, we have $\tr \g^{(1)}=\tr \g^{(0)}\partial_s\g^{(0)}=0$. As a 
result $\g^{(1)}\propto i\sigma_y\t2$. That is, $\g^{(1)}$ does not contribute to the
transport current, so the leading corrections to our results for the superheating field are 
of order $\epsilon^2\sim\kappa^{-2}$.

\end{widetext}


%
\end{document}